
%

\input  harvmac

\def\L{\Lambda}
\def\D{\Delta }
\def\CO{{\cal O}}
\def\CP{{\cal P}}

\def\W{\Omega}
\def\o{\omega}

\def\v{\hat v}

\def\cmp{{\it Commun. Math. Phys.}}
\def\mpl{{\it Mod. Phys. Lett.}}

\def\tt{ \theta}

\def\tr{{\rm Tr}}

\def\p{\partial}

\def\gst{\gamma _{str}}

\def\R{\relax{\rm I\kern-.18em R}}
\font\cmss=cmss10 \font\cmsss=cmss10 at 7pt
\def\cmss{\relax}\def\cmsss{\relax}
\def\Z{\relax\ifmmode\mathchoice
{\hbox{\cmss Z\kern-.4em Z}}{\hbox{\cmss Z\kern-.4em Z}}
{\lower.9pt\hbox{\cmsss Z\kern-.4em Z}}
{\lower1.2pt\hbox{\cmsss Z\kern-.4em Z}}\else{\cmss Z\kern-.4em Z}\fi}
\def\pl{{\it  Phys. Lett.}}
\def\np{{\it Nucl. Phys. }}
\def\cmp{ {\it Comm. Math. Phys.}}
\def\mpl{{\it Mod. Phys. Lett.}}

\def\i{{\rm Im}}

\def\prl{{\it Phys. Rev. Lett.}}

\def\la{\lambda}

\def\wp{\Omega ^{(p)}}
\def\wq{\Omega ^{(q)}}

\Title{\vbox{\baselineskip12pt\hbox{CERN-TH.6834/93 }\hbox{LPTENS 93/7 }}}
{\vbox{\centerline{{Rational  Theories  of 2d Gravity}}
       \vskip2pt\centerline{from the Two-Matrix Model}}}
\centerline{J.-M. Daul   }
\centerline{Laboratoire de Physique Th\'eorique}
\centerline{D\'epartement de Physique de l'Ecole Normale Sup\'erieure}
\centerline{ 24 rue Lhomond, 75231 Paris Cedex 05, France,}

\bigskip\centerline{
V.A. Kazakov \footnote{$^\#$}
{Permanent address: Laboratoire de Physique Th\'eorique,
 D\'epartement de Physique de l'Ecole Normale Sup\'erieure,
 24 rue Lhomond, 75231 Paris Cedex 05, France}
and
I. K. Kostov \footnote{$^\ast $}{Permanent address:
Service de Physique Th\'eorique de Saclay,
 CE-Saclay, F-91191 Gif-sur-Yvette Cedex, France}   }

\centerline{Theoretical Physics Division, CERN}
\centerline{CH-1211 Geneva 23, Switzerland}

\vskip .3in

\baselineskip 8pt
The correspondence claimed by M. Douglas between the multicritical
regimes of the two-matrix model and 2d gravity coupled with $(p,q)$
rational matter field, is worked out explicitly.
We found the minimal $(p,q)$ multicritical potentials $U(X)$ and $V(Y)$,
which are polynomials
of degree  $p$ and $q$, correspondingly. The loop averages    $W(X)$
and $\tilde W(Y)$ are shown to satisfy the Heisenberg relations
$\{ W,X\} =1$ and $\{ \tilde W , Y\} =1$ and essentially coincide with
the canonical momenta $P$ and $Q$.
The operators $X$ and $Y$ create the two kinds of
boundaries in the $(p,q)$ model
related by the duality $(p,q) \leftrightarrow (q,p)$.
Finally, we  present a closed   expression for the two two-loop correlators,
and interpret its scaling limit.

\bigskip
\centerline{Submitted for publication in: {\it Nuclear Physics B}}
\rightline{ }
\vfill\leftline{CERN-TH.6834/93}
\Date{March 1993}

\baselineskip=20pt plus 2pt minus 2pt

\newsec{Introduction}
The theory of non-critical strings, or  2d gravity with  central
 charge of the matter $c \leq 1$, has  made great steps forward
 in the  last years.
  This theory allows  two very  different  formalisms:  the Liouville
 theory \ref\polya{A.M.Polyakov, \pl  B 103 (1981) 207}
   and the two dimensional Regge calculus based on matrix models
\ref\david{F.David, \np B 257 [FS14] (1985)45}, \ref\kazak{V.A.Kazakov,
\pl B 150 (1985) 282}, \ref\kkm{V.A.Kazakov, I.K.Kostov and A.A.Migdal,
  \pl B 157 (1985) 295}, which are in a sense complementary to each other.
Spectacular results
    in calculations of critical exponents  \ref\isingi{V.A.Kazakov,
 \pl A 119 (1986) 140;  D.V.Boulatov and V. Kazakov,
\pl B 186 (1987) 379},
 \ref\km{V.A.Kazakov and A.A.Migdal, \np B  311 (1989) 181},
  \ref\on{I.K.Kostov, {\it Mod. Phys. Lett.} A4 (1989) 217},
 \ref\kpz{V.Knizhnik,
   A.Polyakov and A.Zamolodchikov,  {\it Mod. Phys. Lett.} A3 (1988) 819},
 \ref\davidii{F.David, {\it Mod. Phys. Lett.} A3 (1988) 1651 },
 \ref\dika{J.Distler
 and H.Kawai, \np B 321 (1989) 509}, macroscopic observables
 \ref\kostov{I.K.Kostov, Proceedings of the Carg\`ese Workshop 1990;
\pl 266 B (1991) 42 and 317; \np B 376 (1992) 539},
 \ref\rutgers{G.Moore, N.Seiberg and M.Staudacher,
 \np B 362 (1991) 665}, \ref\goul{M. Goulian, \pl 264B (1991) 292} \
 and correlation functions
 \ref\kutdifr{P.Di Francesco
and D.Kutasov, {\it Nucl.Phys.} 342 (1990) 589},
 \ref\corr{M. Goulian and M. Lee, \prl   66 (1991) 2051},
 \ref\dotsen{V.Dotsenko, Proceedings of the Carg\`ese
meeting 1990} have been accomplished.
   One of the important achievements was a non-perturbative analysis
  of 2d gravity in the framework of the double
scaling  limit \ref\ds{M.Douglas and S.Shenker, \np B 335 (1990),
E.Brezin and V.Kazakov, \pl 236B (1990) 144, D.Gross and
A.A.Migdal, \prl 64 (1990) 127}, which found its most elegant
 realization in the KdV hierarchy of string equations
\ref\kdv{M.Douglas, \pl 238B (1990) 176}.

The generic rational string theory can be constructed as a chain of coupled
random matrices \kdv ,  \ref\sos{I.K.Kostov, Gauge-invariant matrix
 model for the
A-D-E closed strings, \pl 297B (1992) 74}.
However, one can look for a more economic way to do it.
In this paper we address the problem of finding the simplest
 model of (discretized) 2d gravity with matter implementing the whole
 physics of non-critical strings; or, in other words, finding  the
 simplest matrix model containing all known critical
    regimes corresponding to $(p,q)$ conformal matter.

The first attempt to build such a minimal description was made in ref.
\ref\mcrit{V.Kazakov, {\it Mod. Phys. Lett. } A4 (1989) 2125 },
 where various matters coupled
with gravity were generated as multicritical regimes of the one-matrix
model. Since the random surface representation
(Feynman graphs) of this model does not contain any explicit
introduction of matter  (like spins at  the sites  of the random lattice)
this latter appears  as the multicritical excitations of ``pure
gravity'' modes. It soon became clear \ref\stau{M.
Staudacher,  \np B 336 (1990) 349 },
 \ref\dising{E.Brezin, M.Douglas, V.A.Kazakov
and S.Shenker, \pl 237B (1990) 43} that the multicritical
 regimes in the 1-matrix model
 correspond only to the       special case  $(q=2, p=2n-1)$.

 The   next simplest possibility is the two-matrix model first introduced in
\ref\itzub{C.Itzykson and J.-B.Zuber, {\it J. Math. Phys.} 21 (1980) 411 }
 and solved in
the large $N$ limit in \ref\mehta{M.L.Mehta, {\it Comm. Math. Phys.}
79 (1981) 327}. As was pointed out
by M.Douglas \ref\doug{M.Douglas, Proceedings
     of the Carg\`ese Workshop, 1990},
 one can get,
by appropriately adjusting the potentials of both
     matrices,
  the critical regimes
     corresponding to various $(p,q)$-rational string theories,
     which in turn correspond to 2d gravity interacting with matter fields
     having central charge
\eqn\central{  c=1-6 { (p-q)^2 \over p q }. }

Some examples of this conjecture for particular values of $p$ and $q$ were
considered in
\ref\somex{
T. Tada, {\it Phys.Lett.} B259 (1991) 442, \
S. Odake, \pl 269 (1991) 300, \
T. Tada and T. Yamaguchi, \pl 250 (1990) 38,\
}.

Technically, this means      that in the realization of the Heisenberg
algebra $[P,X]=1/N$ in the basis of orthogonal polynomials of the two-matrix
 model     \mehta, \ref\heisen{Chadha, Machoux, M.-L. Mehta, {\it
J. Phys. A: Math. Gen.} 14 (1981) 579 }, the operators $P,X$ are the
 infinite matrices containing
enough diagonals (unlike those in the one-matrix model) to describe
the differential operators of a generalized
 KdV hierarchy of an arbitrary order.
The appearance of   generalized KdV flows in the
two-matrix model was studied  in ref.\ref\toda{
E.Martinec, \cmp 138 (1991) 437;\
S. Ryang, {\it Phys. Rev. } D46 (1992) 1873;\
 A. Marshakov, A. Mironov and A. Morozov, \mpl A7 (1992) 1345;\
 E. Gava and K. Narain, \pl B263 (1991) 213
}.

In this paper, we present the explicit construction of the
 $(p,q)$ critical points
 of the two-matrix model.
The deviation from the critical point is due
 to the shift of the ``cosmological constant''
from  its  critical value \foot{Strictly speaking, in the case
of a non-unitary critical point $p>q+1$ the coupling of the matrix model is
not the cosmological constant. The latter is coupled with the identity
operator,
 which in this case is not the operator of minimal dimension. We will
 discuss this point later.}, while the matter remains critical.
 Consider for example the point $(3,4)$ describing the Ising
model on a random graph     \isingi .
The critical potential means that we are exactly at the spin ordering
phase transition point, and by shifting
the cosmological constant (the overall  coupling in the matrix model)
from its critical value
we make the volume of the system big but finite. The existence of an
 explicit and
relatively simple formula for the $(p,q)$ multicritical potentials
explains  why one
was able to find explicitly the critical ``temperatures'' in many
particular cases, such as Ising models on various types of random graphs.

We start from the  equations of motion, relating the canonical
momenta $P, Q$ to the coordinates $X,Y$,
as they were  derived (but not extensively used) in     \doug . We
wrote them in an operator form,  valid for any $N$ but adjusted to
the double scaling limit. The method looks quite promising for
more complicated matrix models (in \ref\induce{V.A.Kazakov, preprint
CERN-TH.6754/92, November 1992,
 to appear in {\it Nucl.Phys.}B, Proceedings of the Conference
``Lattice-92'', Amsterdam (1992)} it was applied to the so-called
induced gauge theory).

We  extract the critical potentials for  the $(p,q)$ critical regime
from the algebraic form of the equations of motion.
In order to achieve the $(p,q)$ critical point it is sufficient to
take as  the potentials $U(X)$ and $V(Y)$ two polynomials of degree
$p$ and $q$, respectively: this allows for easier, more explicit and
more general formulae than in ref.\ref\deboer{Jan de Boer,\np B336
(1991) 602 }.
For investigating the one-loop averages, we exploit the
Heisenberg algebra for $[P,X]=1/N$,
which appears in the formalism of orthogonal polynomials.
In the large $N$  (classical, planar)
limit, the commutator  reduces  to the  Poisson bracket $\{P,X\}$
of the corresponding
``classical'' functions. The most important observation here, from
 our point of view,
is that the ``KdV-momentum'' $P$ is equal
to the 1-loop correlator
$W(x)=\langle {\tr \over N} {1 \over x-\hat X}\rangle $
(up to some part, independent from the cosmological constant, giving no
contribution
to the Poisson brackets). This is true even before the continuous limit,
and it gives a direct physical meaning to the Heisenberg relation:
$\{W,X\}=1$, allowing the direct calculation
of the macroscopic observable
$W(x)$.

Using all this we were able to rederive in a simple way the results
of     \kostov ,     \rutgers\  for $W(x)$ in the general $(p,q)$ case.
The unitary $(m+1,m)$
series can be realized in two different ways. The first is just the
general  $(p,q)$ solution with $p=m+1, q=m$.
The second way is to look for criticality under the condition
that the
two  potentials $U$ and $V$  are equal.
These two realizations turn to be physically very different.

In the non-symmetric realization the operators $X$ and $Y$
have different scaling dimensions and create two
different kinds of boundary.
Our conjecture is that they correspond to the Dirichlet and Neumann
boundary conditions in the statistical realization of the $(m+1,m)$
model.
The  symmetric realization can be interpreted as a $m$
multicritical regime of
the Ising model on a random graph.
Then the operators $X$ and $Y$ are related by the $\Z_{2}$ symmetry in a
very simple way.

The situation is more complicated for the non-unitary cases;
what we called, in the unitary case, the ``cosmological constant''  $\la$
(the
overall constant in front of the matrix potential) couples there
with an operator of lowest negative
scaling
dimension $\Delta_{-} = {1-|p-q| \over p+q-1}$     \dising .
The genuine  cosmological constant $\Lambda$  is coupled to the unit
operator and measures the  total volume
of the random graph. These two constants are related by
\eqn\lala{ \la = \Lambda ^{{p+q-1 \over 2q}}.}

 In a recent paper  \ref\krich{I.Krichever,
preprint LPTENS-92/18 (May 1992)}  the general solution
of dispersionless Heisenberg equations  is
given in terms of the $\tau$-function of the Whitham hierarchy.
This ``solution'' is not very explicit, but it provides
a general description of the vicinity of the critical point.
Therefore we give a presentation of our results from the point of view of
ref.\krich .

Finally, we found the general expression for all  two-loop correlators
 and investigated their scaling limit,
which coincided with the known ones     \kostov,     \rutgers.
In the scaling limit we observed a remarkable factorization property,
inspiring a topological field theory interpretation.
It generalizes the formula obtained
in \ref\make{J.Ambjorn, A.Krzyvicki and Yu.M.Makeenko, \pl B251 (1990) 517}
 in the
case of the one-matrix model.

This paper is organized as follows:

In     section 2 we formulate the model, and using the orthogonal
polynomials formalism, derive the general equations
of motion and their planar limit.

In section 3 we introduce the Heisenberg algebra for the matrix
eigenvalue operator and its momentum, obtain their
planar limit in terms of the Poisson brackets and establish the
equality between the one-loop average and the Heisenberg momentum.

In section 4  an explicit formula for the $(p,q)$ multicritical
potentials in the two-matrix model will be derived.

In section 5 we  demonstrate how the equations of motion work on
some examples of unitary as well as non-unitary theories.

In section 6, using the scaling limit of the classical Heisenberg
relation (dispersionless KdV) we reproduce the known results
for the different  loop correlators.

In section 7 we discuss the vicinity of the critical point,
in view of our interpretation of the loop average as the classical
canonical momentum. We   review the
operator content of the theory.

In section 8 we interpret  the  two kinds of boundaries in the model
as the Dirichlet boundaries in the two alternative standard realizations
of the $(p,q)$ critical point with $q-1$ and $p-1$ spins.

In section 9 the two-loop correlator is calculated explicitly in the
two-matrix model, and its scaling limit is found to be consistent
with the known results.

In section 10 we summarize the results for the $\Z_{2}$-symmetric
two-matrix model describing the unitary series $(m+1,m)$.

\newsec{ Orthogonal polynomials in the two-matrix model with
 two arbitrary potentials}

The partition function for the two-matrix model is defined as
an integral over two $N\times N$ hermitean matrices $X$ and $Y$
\eqn\partf{Z = \int d^{N^2}X \ d^{N^2}Y \
e^{ \beta \tr (-U(X)-V(Y)+XY)}}

\noindent where $U$ and $V$ are arbitrary polynomial (or even non-polynomial)
potentials:
\eqn\pot{U(X)= \sum_{n=1}^{p} {g_{n} \over n} X^{n},
V(Y)= \sum_{n=1}^{q} {h_{n} \over n} Y^{n}.}
The linear change of variables $X\to A_{1}X+A_{2},Y\to A_{3}Y
 +A_{4}$  has a trivial Jacobian and preserves the form of the
potential. Therefore the problem depends on $p+q-3$ relevant directions
in the space of parameters $h_{1},...,h_{q},g_{1}, ... , g_{p}, \beta$
in  \pot .

As was stated in \isingi , the Feynman graph expansion of this model,
with respect to the non-quadratic terms in the potentials of \partf\ , can
be interpreted as the sum over graphs of partition functions of Ising
spins placed in the vertices of these graphs. Since the Ising model in
two dimensions describes fermions, and the sum over Feynman graphs can
be interpreted as the functional integral over the two dimensional metric,
it can be concluded that we deal with fermions interacting with 2d
gravity. As is well known, the $1/N$ expansion allows the classification of
these partition functions according to genus, so that the total
partition function can be viewed as the functional integral of the
fermionic string field theory with $1/N$ as a string coupling. So
in the limit $ N \rightarrow \infty $ we are left with only planar 2d
manifolds.

On the other hand, the representation of 2d manifolds in terms of
planar Feynman graphs depends on the couplings in the potentials
\pot. One can perform the discretization by choosing either
$\phi^3$-  or any $\phi^n$-potential,
 or even any mixture of vertices of different
orders. As in the one-matrix model \mcrit , one can adjust the
corresponding couplings so as to get different multicritical
behaviours, corresponding, in the two-matrix case, to all possible
$(p,q)$ minimal models coupled with 2d gravity. One can say that
(unlike the case of SOS matter \sos )
the $(p,q)$ matter is produced,  in the two-matrix model, as the collective
excitations of
the $Z_2$ (Ising) degrees of freedom and those of the discretized
metric of the world-sheet.

Let us start with briefly reviewing the method of diagonalization due to
orthogonal polynomials.
If we integrate with respect to the angular degrees of freedom,
 \partf\  becomes an integral over the  $2N$ eigenvalues
$x_{1},...,x_{N},y_{1},...,y_{N}$:
\eqn\diag{Z=\int \prod_{i=1}^{N} dx_{i}dy_{i}\D (x)\D (y)
e^{\beta  \sum
_{i=1}^{N}  (-U(x_{i}) -V(y_{i}) +x_{i}y_{i})}.}
We introduce the orthogonal polynomials $|n\rangle \ =\zeta_n(x)$ and
$\langle n|\ =\eta_n(y)$
by the orthonormality relation
\eqn\i{\langle m |n \rangle  = \int dx \ dy
  e^{\beta (-U(x)-V(y)+xy)} \zeta _{m}(x)\eta_{n}(y) =  \delta_{n,m} .  }
 We denote
the  matrix elements:
\eqn\ii{X_{mn}= \langle m | x | n \rangle ,\
\        Y_{mn}= \langle m | y | n \rangle ,
  \  P_{mn}= \langle m | {\p \over \p x}| n \rangle ,
  \  Q_{mn}= \langle m | {\p \over \p y}| n \rangle .}
The following ``equations
of motion'' are easily obtained by doing an integration by parts:
\eqn\iii{{1 \over \beta}P_{mn}= \langle m|U'(x)| n \rangle - Y_{mn},
\ \ \ \ \ {1 \over \beta} Q_{mn}= \langle m|V'(y)| n \rangle - X_{mn}.}
At this point it is useful, with a view to the later introduction of
differential operators in the double scaling limit,
to change the notations for the indices
of the matrix elements:
\eqn\indi{ X_k(n)=X_{n-k,n} ,     \ \       Y_k(n)=Y_{n,n-k} . }
Now $n$ marks the position of the matrix element on the diagonal, and
$k$ is its deviation from it. Then the action of the operators
$X,Y$ on the orthogonal polynomial basis is described by the formulae
\eqn\iv{X | n \rangle = \sum _{k =-1}^{q-1} X_{k}(n)
|n-k \rangle;
\ \ \langle n |Y = \sum _{k= -1}^{p-1} Y_{k}(n) \langle n-k|.}

To prepare for both the continuum and planar limit, introduce the
continuous    variables
\eqn\v{  t = n/\beta,\ \  \lambda = N/\beta , }
     the operator $\hat n$ whose eigenvalues are $n$
(i.e. $ \hat n| n\rangle = t|n \rangle, t = n/\beta$),
and the conjugate  coordinate
$\o$, as it was prescribed in ref.
\ref\gmm{D. Gross and A.A. Migdal, \np B 340 (1990) 333}.
 Then $\hat n$ is represented by the differential operator
$\hat n= - (1/\beta)  d/d \o $ , where $ \beta={N \over \lambda} $.
The constant $1/\beta$ plays the role of the Planck constant in this
formalism. The operators  $X$  and $P$  are  represented by the functions
\eqn\vi{
X(\o , t)
=\sum  _{k = -1}^{q-1}  e^{k\o} X_{k}(n),
\ P(\o,t) = \sum _{k = 1} ^{(q-1)(p-1)}e^{k\o} P_{k}(n);
\ \ \ t= n/\beta.}
Similarly one defines the functions $Y (\o , t), Q (\o , t)$ representing the
operators $Y$ and $Q$
\eqn\vij{Y(\o , t)
= \sum  _{k = -1}^{p-1}  e^{k\o}  Y_{k}(n),
\ Q(\o,t) = \sum _{k = 1}^{(p-1)(q-1)}  e^{k\o} Q_{k}(n) ;
\ \ \ t= n/\beta.}
It is always possible to normalize the polynomials \i\ so that
\eqn\nnnor{ Y_{-1}=1, Q_{1}=t.}
 Then the coefficient $X_{-1}$ is
determined as a function of $t=\la $:
\eqn\ccco{X_{-1}(n)= R(t), \ \ P_{1}(n)= {t \over R(t)}\ ,}
where $R(t)$ is related to the susceptibility
\eqn\ssu{R(\lambda)\sim {\p ^{2} \over \p \lambda ^{2}}\log Z.}
Note that for any fixed highest power of the polynomial potentials
\pot , we have fixed highest powers with respect to $z=e^\o$ in \vi\ and
\vij.

Now eqs. \iii\ read
\eqn\vii{\eqalign{
  P(\o , t)&=
U^{'}\bigg(X\big(\o,\ t-{1 \over \beta} { \p \over \p \o } \big) \bigg) \cdot 1
- Y\bigg( - \o +{1 \over \beta}{ \p \over \p t } ,\ t \bigg) \cdot 1\ ;\cr
 Q(\o , t)&=
V^{'}\bigg( Y\big(\o, t -{1 \over \beta}{ \p \over  \p \o } \big)  \bigg)
\cdot 1- X\bigg(- \o +{1 \over \beta} { \p \over \p t }, \ t \bigg)  \cdot 1 \
.\cr}}
The compatibility of eqs. \vii\ with the condition that $P(\o,t)$ contains
only positive powers of $e^{\o}$ lead, in the limit $N \to \infty$,
 to a triangular system of algebraic equations for the coefficients
$X_{k}$ and $Y_{k}$ in the expansions \vi\  and \vij .

The operator equations \vii \ are exact and valid for any  $N$. They are
very useful to go to the planar limit or to the double scaling limit.
In the planar limit, we have to drop all the derivatives in the arguments
of the functions $X,Y$ in \vii , and we arrive to the following
system of (algebraic) equations:

\eqn\viik{\eqalign{
 P(\o ,t)&= U^{'}\big(X(\o, t)\big) - Y( - \o ,t) \cr
 Q(\o , t)&= V^{'}\big(Y(\o,t)\big) - X( - \o ,t) .}}
Again, the condition that $P,Q$ have only positive powers of $e^\o$ is
sufficient to find all the functions $P,Q,X,Y$.

In the  symmetric case  $U=V$   the two equations \viik\ reduce to
a single equation for $P(\o , t)=Q(\o,t)$
 and $X(\o , t)=Y(\o , t)$
\eqn\ooio{ P(\o ,t)= U'(X(\o , t))- X(-\o , t),}
where, choosing the symmetric normalization of the polynomials,
\eqn\huy{ P={t \over \sqrt{ R}}e^{-\o}+..., X=\sqrt{R}\, e^{\o}+...}
This equation allows the coefficients $X_{k}$ and $P_{k}$ to be expressed in
terms of the potential $U$. Once the potential is fixed, the first
coefficient $R$ is found as a function of $t=\lambda$. For the $(p,q)$-critical
potential it behaves as $R \sim \lambda ^{2/p+q-1} $ + const.

The explicit expression of $X$ and $P$ can be found more easily by means of the
canonical commutation relations, to be studied in the next section.

\newsec{Heisenberg algebra, its classical limit
and  loop correlators in the planar limit}
It is obvious from the definitions \ii  of the operators $X,Y,P,Q$ that
they obey the Heisenberg commutation relations:
\eqn\heis{[P,X]={1 \over N} ,\  [Q,Y]= {1 \over N}.}

In the classical limit $N \to \infty$ the commutator in eq. \heis\
 is replaced by the
Poisson bracket
\eqn\pb{\{ P,X\}= {\p P \over \p t}{\p X \over \p \o}
-  {\p P \over \p \o}{\p X \over \p t}=1
.}

Let us now consider the loop average
\eqn\loop{W(x,\lambda) = \bigg\langle {\tr \over N}{1 \over
x-\hat{X}}\bigg\rangle}
as a  function of $\o$ and $\la$ through  $x=X(\omega,\lambda)$:
 $W = W(X(\o, \lambda), \lambda) $. We will show that this function has the
same Poisson bracket
with the coordinate $X(\o,\lambda)$ as the momentum $P(\o , \la)$.
 By definition of the average in the basis of orthogonal polynomials:
\eqn\wll{ {\p \over \p \lambda } W_1(x, \lambda)=
 {1 \over 2 \pi i} \oint d \o {1 \over x- X(\o,\lambda)}=
\Bigg({1 \over  \p_{\o}
X|_{\lambda}}\Bigg)_{X(\o,\la)=x}
={\p \o \over \p X}\Big| _{\la}
.}
On the other hand,
\eqn\wxw{
{\p W \over \p \la }\Big|_{X}={\p W \over \p \la }\Big|_{\o}-
{\p W \over \p \o}\Big|_{\la} {\p_{\la}X|_{\o} \over \p _{\o}X|_{\la}}
={\p \o \over \p X} \Big|_{\la}\{ W, X\}\ ,}
which implies
\eqn\wx{\{W(\o,\lambda),X(\o,\lambda)\} = 1 .  }
Similarly, for
\eqn\loopp{\tilde W(y,\lambda) =
\bigg\langle {\tr \over N}{1 \over y-\hat{Y}}\bigg\rangle }
we obtain
\eqn\wy{\{\tilde W(\o,\lambda),Y(\o,\lambda)\} = 1 . }
Equations \wx\ and \wy\ mean that
\eqn\wpp{ W(\o,\lambda)= P(\o,\lambda) + f(x)  }
\eqn\wpp{ \tilde W(\o,\lambda)= Q(\o,\lambda) + g(x) \,, }
where the functions $f(x)$ and $g(x)$ do not depend on $\lambda$.
Indeed, the explicit expression for $P$ as a Chebyshev polynomial
\kutdifr , \ref\zinb{B. Eynard and J. Zinn-Justin, Preprint SPhT/92-163}
coincides, as a function of $X$,
 with the expressions of the Wilson loop  found in \kostov ,
\ref\nonr{I.K. Kostov, \pl 266 (1991) 42}.

Thus the classical Heisenberg algebra \wx\ and \wy\ is now
formulated in terms of the basic macroscopic observables, the loop
averages.

Equation \wll\ , which can be written as
\eqn\llop{{\p P \over \p \la}\Big|_{X}=
{\p \o \over \p X }\Big|_{\lambda}}
implies that there exists a ``potential''  $S$ such that
\eqn\pppd{P={\p S \over \p X}, \ \o = {\p S \over \p \lambda}.}
The differential of this function is
\eqn\diffs{dS = PdX-\o  d\lambda .}
  The function $S(X,t)$ makes sense of  the action of the Hamiltonian system,
as a function of
 the coordinate $X$ and
the time $ t= \lambda$ along the classical trajectories.

It follows from \wpp\ , \loop\
 and \pppd\ that the function $S$ may be obtained
as the average
\eqn\ioiid{
S(x,\la)= \bigg\langle {\tr \over N} \log (x - \hat X) \bigg\rangle .}

Similarly, we could start with the $y$ variable and the action
\eqn\ioiidd{
\tilde S(y,\la)= \bigg\langle {\tr \over N} \log (y - \hat Y) \bigg\rangle \ ,}
which is, in a certain sense, dual to \ioiid .
Indeed, by the equations of motion \viik\
  the canonical pair of variables $(Q,Y)$ is obtained from $(X,P)$ by
the involution $\o \to -\o$.

The relations of the type  \wx\
found in this section are, of course, valid far
beyond the two-matrix model, since the proof \wll\ and \wxw\ is general
enough. It is certainly true for the one-matrix model,   for the
multi-matrix chain and for any matrix model where the orthogonal
polynomial formalism is available. One can even contemplate that
these formulae are not specific to the orthogonal polynomial
representation, and may be, some general approach to the string
theories could be based on it.

\newsec{Explicit $(p,q)$-multicritical potentials of the two-matrix model}
The first step when investigating the $(p,q)$ critical behaviour
of the two-matrix model is to find a set of critical couplings $g_k,
h_k$. Near such a point in the space of couplings the size of the planar graphs
tends to infinity and the corresponding $(p,q)$ ``matter'' exhibits
long-range fluctuations.
There are infinitely many ways to achieve
 the $(p,q)$ critical regime by varying the coupling constants;
 we choose the most economic  option, which corresponds to potentials
$U$ and $V$ of minimal power $p$ and $q$.
Even this restriction does not fix the multicritical potential
uniquely.  We take advantage of the residual freedom to
fix  potentials  with all coefficients rational.

  At the critical point $\la = \la ^*$ one expects the following
 singular
behaviour of $P$ and $X$
\eqn\ccr{X - X^{*} \sim \o^{q}, \ \
P - P^{*} \sim (X-X^{*})^{p/q} =\o^{p} \ \ \ \ (e^{\o}=z) }
Inserting this in \viik\ , we see that the singular behaviour of $Q$ and $Y$
can only be
\eqn\ssusi{Y -Y^{*} \sim \o^{p}, \ \ \
Q - Q^{*} \sim  \o ^{q}= (Y-Y^{*}) ^{q/p}}
The  singular behaviour \ccr\ and \ssusi\ is sufficient to fix the form of the
functions \vi\ and \vij\ (we normalize $X$ so that $R=1$)
\eqn\vxc{X_{*}(z)= {(1-z)^{q} \over z},\ \ \
Y_{*}(z)= {(1-z)^{p} \over z}}
The formal solution of \viik\ in terms of contour integration
around $z=0$  gives the $(p,q)$ multicritical potentials
\eqn\jky{\eqalign{
U'_{p,q}(\xi)&= - {1 \over 2\pi i} \oint {Y_{*}(1/z)
\over \xi -X_{*} (z)}  dX_{*}(z)
 \cr
V'_{p,q}(\eta)&=-  {1 \over 2\pi i} \oint { X_{*}(1/z)
\over \eta -Y_{*}(z)}
 dY_{*}(z) \cr
}}
Here the contour of integration excludes all the poles except that at $z=0$.

 Note that only the negative powers of $z$ are relevant for the definition
of the potential from the given $X_*(z)$ and $Y_*(z)$, therefore $P(z)$ and
$Q(z)$ do not enter the final expressions \jky\ . It is easy to see that the
formulae  \jky  indeed give the polynomials of the p-th and q-th degree,
respectively.

The unitary series of multicritical points  $p=m+1, q=m$ can be
achieved also for equal potentials  $U=V$ .
If we take $X_{*}(z)=Y_{*} (z)= (1-z)^{m}/z$ , then the $m$-critical
potential given by \jky\ will describe the point $(m,m+1)$ for $m=3,4,... $\ .
For example, the case $m=3$ will give the
Ising model on a 3-coordinate random lattice (with tadpole and self-energy
 subgraphs included).

The multicritical potentials for a few lowest $p$ and $q$, as well as for the
 symmetric case for a few lowest $m$, are presented in Appendix A.
The same calculation can be carried out in the frequently used case of
the potentials containing only even powers of matrices $X$ and $Y$. The
details are presented in section $10$. In Appendix A the results
for the few first $m$'s in the $Z_2$-symmetric case $U=V$ are presented.

A remarkable feature of this symmetric realization is that for every even
$m$ the corresponding multicritical potential is  stable
(the coefficient in front of the highest power is positive),
which provides a proper and quite economical way to non-perturbatively
define
the double scaling limit of the model.

\newsec{ Some examples}

Should we try to find the classical functions $X(\o,\la), Y, P, Q$ when
given the
 potentials,
we would have to solve a triangular system of algebraic equations, with degrees
 as high as $p$.
On the other hand, expressing the potentials in terms of the
 coefficients of $X$
 is fairly easy,
as it involves only linear equations. Having recognized the critical form of
 $X,Y,\ldots $(which are polynomials, up to some $ 1/ z$
 term), we find critical potentials,
and we can then try
to move away from the critical point in the coupling constants space and
 see how
 $X,Y,\ldots$
have to be shifted : this can be done perturbatively, for any degree. In this
 way, we may
access the scaling behaviour of our macroscopic observables, and check whether
 it satisfies
the Heisenberg relation.

We shall first consider the symmetric case : the two coordinates are equal, and
 the one-loop
correlator is given by the classical momentum (up to some function of
 $X$).
Here, we reproduce the $(m,m+1)$ unitary case, with :
\eqn\casunit{ \left\{ \eqalign { &X_*(z)\sim (1-z)^m \cr
 &W_*(z)\sim (1-z)^{m+1} \cr }\right. }
The corresponding critical behaviour appears when we shift $\la$ away from its
 critical value:
this is a particular move in the space of parameters, which involves the
 identity operator
of the world-sheet in this unitary case, as will be explained later. Here, we
 expect
the susceptibility to behave as $(\la - \la _c)^{1/m}$.

The simplest situation is the Ising model on a triangulated lattice ($m=3$):
we
 may solve
the algebraic equations \viik\ explicitly and extract the scaling behaviour of
 $X$.
If we substitute the critical form of $ X_*(z)=\displaystyle{(1-z)^m / z}$
\ in
\eqn\potinconnu{P(z,t)\ =\ U'\big(X(z,t)\big) - X\bigg(\displaystyle{1\over
 z},t\bigg) }
then claim that $U'(\varphi)=\alpha +\beta \,\varphi + \gamma \varphi^2$, and
 identify different
powers of $z$ ( we know that $P_*(z)=t_*\,z\ +\ {\rm higher\,powers\,of\,}z$ )
 we may
identify the critical potential:
 \eqn\potident{U'(\varphi)=3-3\,\varphi-\varphi^2}
and the critical value $t_* \equiv \la _* =10$.
Now, we try to find $X$ for any $t$ :
\eqn\xincon{X(z,t)=\displaystyle{\sqrt R \over z}+a+b\,z+c\,z^2}
again by identifying powers of $z$ in \potinconnu\ and using
 $P(z,t)={t\over \sqrt R}
z+\ldots$.
This leads to the following system :
\eqn\systabc{\left\{ \eqalign {
0&=R+c \cr 0&=3\sqrt{R}+2\,a\sqrt{R}+b \cr
0&=3-4\,a-a^2-2\,b\sqrt{R} \cr
\displaystyle{t\over \sqrt{R}}&=-\sqrt{R}-3\,b-2\,c\sqrt{R}-2\,a\,b \cr }
 \right. }
This way, we can express $c=-R$, eliminate $b$ so as to get a
quadratic equation
 for $a$
with coefficients involving $R$, solve it and keep the only branch that
 corresponds to the
critical value $a_*=-3$.
We obtain:
\eqn\solabc{ \eqalign{a&=2(R-1)-\sqrt{4\,R^2-2\,R+7} \cr
b&=-3\sqrt{R}-2\,a\,\sqrt{R} \cr
c&=-R \cr }}
where $R$ is given through
\eqn\corde {\la \equiv
t=32\,R^3-14\,R^2+28\,R-4\,R(4\,R-1)\sqrt{4\,R^2-2\,R+7} }
Finally, we obtain (implicitly) an exact expression for $X$, whose scaling
 behaviour is found
by letting $\la \to \la_c$ and $z-1 \sim (\la-\la_c)\,^{{1 \over 2m}} $.
We find the third Chebyshev polynomial (up to some arbitrary scaling on $X,z$ )
 which
expresses $\cosh (3\theta )$ in terms of $\cosh (\theta )$ (see the
next section).

Indeed, if we consider the generic scaling limit:
\eqn\echgen{\eqalign{ &X(\la,z)\sim (\la-\la_c) ^{{1 \over 2}}
 \xi\Bigg( {(z-1)\over  (\la-\la_c) ^{{1 \over 2m}}} \Bigg) \cr
 &P(\la,z)\sim (\la-\la_c)^{{m+1 \over 2m}}
\  \pi \Bigg( {(z-1)\over
  (\la-\la_c) ^{{1 \over 2m}}} \Bigg) \cr}}
with $\xi$ and $\pi$ polynomials
 of degrees $m$ and $m+1$ respectively,
  and if we express the
 canonical Poisson
bracket in terms of these scaling functions, we find
\foot{This equation was also obtained in \zinb\  }
 \eqn\eqfdl{m\, \xi \pi ' \ -\ (m+1)\,\xi ' \pi\, =\ 1. }
If $m=3$ we find that the Chebyshev polynomials do satisfy this relation.
 In any case, an obvious computation shows that hyperbolic cosines
will always satisfy this scaling relation (compare next section).
Will it always be the only polynomial solution?

Let us consider the $m=4$ case ( critical potentials, obtained in the same way
are given in the Appendix). Again, we try to find the coefficients of
 successive
 powers
of $z$ in the expression of $X(z,t)$;
the equivalent to system \systabc\ involves equations of higher degree, which
 allows for
more than one solution corresponding to the expected form for $X_*$: one has
to choose between different branches of solutions. Among the two possible
 branches, only one
exhibits a proper critical behaviour of the string susceptibility $R \sim
 (\la-\la_c)^{1/4}$.
And the corresponding coordinate function behaves, in the critical vicinity,
 as  the fourth Chebyshev polynomial.
Let us note that  the critical point cannot be reached by increasing the
 cosmological constant from zero
towards its critical value because there is a pure gravity ($m=2$)
singularity on the way (for real susceptibilities).

For $m=5$ the algebra becomes a bit more intricate, and among the three
 possible branches of
solutions, two exhibit a fifth-order critical behaviour and have to be
 considered. One of these
leads to the fifth Chebyshev polynomial for the scaling behaviour of $X$, while
 the other
solution is not reducible to that form under rescalings of $X$ and $z$; it
 corresponds to
\eqn\mcinq{\xi (T)=T^5+5\, T^3+15\, T .}
If we come back to the differential equation involving $\xi$ and $\pi$ and look
 for all its polynomial
solutions ( degrees 5 and 6 ) we shall find Chebyshev polynomials and another
 solution:
\eqn\exotic{\eqalign {\xi (T)&=T^5+5\, T^3+15 \, T \cr
            \pi (T)&=T^6+6 \, T^4+21\, T^2+14 \cr }}
Its interpretation is not yet clear.
It might correspond to another scaling behaviour of the matrix model
around the same critical point.
In any case, although $m$ is odd here, we again find maxima in the plot $\la$
versus $R$ which seem to hide the singularity of fifth order; again, the
 critical point has to be
reached in a very particular way.

Having noticed the existence of this exotic solution, we may come back to the
 $m=4$ case, look for all
possible polynomial solutions and discover that \eqn\mquat{\eqalign {\xi
(T)&=T^4+4
 \cr
\pi (T)&=T^5+5\, T \cr } }
but who has to be  excluded as it
corresponds to the wrong scaling $R \sim  (\la - \la _c)^{1/2} $.
It seems that all possible $\xi ,\pi$ solutions are realized as different
 solutions of the algebraic
equations associated with the matrix model.

Let us now turn to the non-symmetric case $U\ne V$.
As we shall see in the next section, it contains
 all $(p,q)$ critical points
 (the corresponding multicritical potentials are
given in the Appendix).

Now, the (3,4) case reveals two branches of solution: the usual Chebyshev
 polynomials, and
\eqn\cheb{\eqalign  {X &\sim T^4-T \cr
             P &\sim 4\, T^3-3 \cr } }
which gives the proper Poisson bracket ( exotic solution, non-existing  in the
 symmetric case) and the right
critical exponent for $R$.

We also considered the (3,5) case, and found two possible scaling behaviours,
 associated with the two
possible solutions for the Poisson bracket equation in the scaling regime.
Here  Chebyshev polynomials are no longer possible solutions  (as will be
 explained later)
but instead:
\eqn\asymetr{\eqalign {X &= {\displaystyle 1\over 3}\, T^5+5\, T^3\pm 5\,
 T^2+15\, T\pm 30 \cr
            Y &= {\displaystyle 2\over 3}\, T^3 +6\, T\mp 6 \cr }}
these being two solutions of \pb \  ( to verify this, do not forget to change
 signs,
 as $P(\o)$ scales
as $Y(-\o)$ ).

A solution for $P,Q,X,Y$ in the unitary case $(p=q+1)$
will be presented in the next section. The complete classification of
scaling behaviours is far from being accomplished.

\newsec{ Explicit solution of the Heisenberg relation
near the  $(p,q)$ critical point}

Once the scaling laws are established, it is more convenient to work
with the differential form of the equations of motion \pb .
Let us use the parametrization
\eqn\ooogp{ \o = 2\mu \cosh (\theta)}
 Near the $(p,q)$ multicritical point, $X \sim Q \sim \mu ^{q}$ and
$Y \sim P \sim \mu ^{p}$.
We know by eq. \pb\
that $\la \sim \mu ^{p+q-1}$.
We introduce two complete systems of polynomials in $\mu$ and $\o$ ,
homogeneous of degree $p$ and $q$
\eqn\pppoi{\eqalign{
\wp _{k}&= 2\mu^{p}\cosh (k\tt), \ \ \ k=0,1, ... , p \cr
 \wq _{s}&=2 \mu ^{q} \cosh (s \tt), \ \ \ s=0,1, ... , q \cr}}
These polynomials have Poisson brackets
\eqn\ppb{ \{\wp_{k},\wq_{s}\} = {\mu ^{p+q-1}\over \p \la /\p \mu}
\Bigg[ (ps+qk){\sinh [(s-k)\tt] \over \sinh \tt}
+ (ps-qk){\sinh [(s+k)\tt] \over \sinh \tt}\Bigg]}
Thus, expanding the functions $X$ and $P$ as linear combinations of \pppoi\
\eqn\chee{P=\sum_{k=0}^{p}A_{k}\wp_{k},\ \ X=\sum_{s=0}^{q}B_{s}\wq_{s}}
\eqn\chee{Y=\sum_{k=0}^{p}(-)^{k}
A_{k}\wp_{k},\ \ Q=\sum_{s=0}^{q}(-)^{s}B_{s}\wq_{s}}
we find from \pb\ a linear system of equations for the coefficients.

This method works particularly well in the unitary case $q=m, p=m+1$.
Then we have simply
\eqn\podo{\eqalign{
P&=\wp_{p}=2\mu^{m+1}\cosh [(m+1)\theta]
,\cr X&=\wq_{q}=\mu^{m}\cosh (m \theta),\cr
 \la & = {pq \over p+q-1}\mu^{p+q-1} =(m+1)\mu^{2m}\cr }
}


\newsec{The vicinity of the critical point}

The interpretation of the momentum as the Wilson loop variable
is valid also outside the critical point.
In this section we will sketch the situation when the potentials $U$ and $V$
represent an arbitrary perturbation of a critical potential.
For more details see the original paper by Krichever \krich .

 Let us parametrize the vicinity of the critical point
by the coupling constants $t_{1}=\lambda, t_{2}, t_{3}, ...$
Then the
operator  $X$ will be represented in the scaling limit
by a general polynomial
 of degree $q$.
As before we introduce
a ``potential'' $ S(\o , t_{1}, t_{2}, ...)$ which in our model has the
meaning of the partition function of a disk without marked point
$\big[ $eq. \ioiid $\big] $.
The function $S(X,t_{1},...) $ can be thought of as the action of a Hamiltonian
system depending on the co-ordinate $X$ and the ``times'' $t_{m},
n=1,2,...$ The total differential of the action is
\eqn\tdrc{dS=PdX -\sum_{n}H_{n}dt_{n}}
where the momentum
\eqn\julp{P={\p S \over \p X}}
is the Wilson loop average
and $H_{n}$ are the Hamiltonians corresponding to the times $t_{m}$.
By the definition  \tdrc\ :
\eqn\huy{H_{n}= - {\p S \over \p t_{n}}\Bigg|_{X}}
 From the commutation of the second derivatives it follows that
\eqn\commmu{{\p P \over \p t_{k}}\Big|_{X}=
-{\p H_{k} \over \p X}, k=1,2,...,q-1}
The relation \wll\ is a particular case of \commmu\ , with $k=1$.
 Let us mention that  dispersionless
KP flows are determined by the Hamiltonian equations of motion
\eqn\tvfyr{{\p P \over \p t_{n}}=\{ P, H_{n} \}, \ \ {\p X \over \p t_{n}}
=\{ X,H_{m} \}}
Each coupling constant $t_{k}$ is coupled with the corresponding scaling
operator $\CO_{k}$.
Equation \tvfyr\ implies the zero-curvature condition
\eqn\zeer{{\p H_{k} \over \p t_{n}}- {\p H_{n} \over \p t_{k}}
+\{ H_{k},H_{n}\}=0}
In the construction of Krichever \krich , the starting point was
 the zero-curvature condition   \zeer\  which is equivalent to
\eqn\gdgdg{\p_m H_n |_X = \p _n H_m |X
}
and thus implies the existence of the action \
such that $H_n=\p_n S|_X$.

Let us consider first  the unitary case $p-m+1, q=m$.
The function $H_{n}$ can be thought of as the partition function of the
disk with a puncture on the world-sheet where the
$\CO_{n}$ operator is placed. For $n=1,2,...,m-1$ this is the $n^{th}$
 order parameter of the
model. The operator $\CP = \CO_{1}$ is the puncture operator corresponding to
the identity operator in the conformal theory $(m,m+1)$.
 The dimension of the time $t_{n}$ is that of the power $\mu ^{
2m+1-n}$, near the point $t_{n}=\delta_{n,1}\mu^{2m}$.
After taking the derivative in $X$ in order to create a marked point on the
boundary, according to \tvfyr\ one obtains
the loop-point correlators that were calculated in
\ref\adek{I. Kostov, \np B  326 (1989)583}\
 for $t_{k}=\mu^{2m} \delta_{k,1}$:
\eqn\iuo{{\p \over \p X} H_{k}(X,t_{1}=\la ,t_{2},...)=   \mu ^{k-m}
{\sinh (k\theta) \over \sinh (m \theta)} \sim {\p \over \p \mu}
(\mu^{m+k} \cosh [(m+k)\theta] )}
Integrating this formula, we find for $t_{n}=\mu^{2m} \delta_{n,1}$:
\eqn\iuy{H_{n} = \mu^{n} \cosh (n\theta)}
In particular, $H_{1}=\o$ (this is true for arbitrary  $t_{k}$).
If we interpret the quantity \commmu\  in terms of an operator $\CO_{m}$
acting on a disk with a puncture, then it should be proportional to
$\lambda^{1-\gst /2 -(1-\Delta_{m})}=\lambda^{1+1/2n -(1-(m-1)/2n)}$
and it follows that the gravitational dimension of the operator $\CO_{k}$
coupled with the ``time'' $t_{k}$ is
\eqn\dimm{\Delta_{k}={k-1 \over 2m}}
Therefore, the operators $\CO_{k}, \ k=1,2,...,m$ are the order parameters
in the $(m,m+1)$ conformal field theory coupled with gravity. The
operator $\CO_{m+1}$ can be identified with the boundary operator
$\p /\p X$ marking a point at the boundary of the loop.
Note that the functions \iuo\ have been  interpreted
in \rutgers\ as the wave-functions of the   on-shell states
of the closed string. They appear as leg factors in the multi-loop
string amplitudes. In particular, the derivative of the two-loop
correlator in the two-matrix model will be expressed as a sum of
products of these functions.

In the general $(p,q)$ case, the coupling $\la = t_{1}$ no longer corresponds
to the puncture (identity) operator, but to the operator with the
minimal (negative) dimension in the model.
This leads, as we have seen in the previous section,
to more complicated expressions for $P$ and $X$.

There exists, however, a direction in which the solution is just
as simple as that in the unitary case. It is along the coupling
constant $\L = t_{p-q}$ corresponding to the identity operator $\CO_{p-q}$.
It seems to us that this is the most natural definition of the non-unitary
critical points. It appears when the $(p,q)$ critical points
 are constructed as
 integrable ADE models on a fluctuating lattice
\adek ,  \ref\dts{I. Kostov, \np B  376 (1992) 539} .
Then the operators with negative dimensions
do not appear by construction and the cosmological constant is coupled with the
volume of the world-sheet.

Along the direction $t_{k}=\delta_{k, p-q}\L$ in the space of the coupling
constants, the potential depends explicitly on the coupling $t=\la$ and
 the quasi-classical limit of \heis\ is no longer given by the
 Poisson bracket relation \pb. Instead, the classical functions
$P$ and $X$
 satisfy the new  Poisson bracket condition
\eqn\hiuihj{ \{ P , X  \}_{\Lambda,\W}
 \equiv {\p P \over \p \L} {\p X \over \p \W}
- {\p X \over \p \L}{ \p P \over \p \W}= 1}
where the new variables $\W$ and $\Lambda$ are related to the old ones
by a non-canonical transformation
\eqn\nonvac{ \W= \mu ^{p-q}\cosh
 [(p-q)\theta], \Lambda = \mu^{2q}}

Equation \hiuihj\ follows from the explicit solution for the loop average
in the   $(p,q)$ critical point of an ADE  model on a random lattice
\dts
\eqn\zzfdg{P(\o , \L)=\mu ^{p}\cosh (p \theta), \ X(\o , \L)=
\mu ^{q} \cosh (q \theta);
\ \ \ \L=\mu ^{2q}, \ \o = \mu \cosh \theta}


\newsec{Boundaries and boundary operators}
In this section we will give an interpretation of the loop operators in
the two-matrix model in terms of statistical systems on
random surfaces.

For a string physicist  $ \tr (x-\hat X)^{-1}$ represents an
operator creating a hole in the world-sheet of the string.
In the two-matrix model (with two different potentials) there are {\it two}
different loop operators
\eqn\llopi{W(x)= {\tr \over N} {1 \over x- \hat X}  ,
\ \ \ \ \tilde W(x)= {\tr \over N} {1 \over y- \hat Y}}
These two operators have different dimensions; their  mean values
read (in the unitary case $p=q+1$)
\eqn\hjk{\eqalign{
W(x)&=(x+\sqrt{x^{2}-\la})^{p/q} +(x-\sqrt{x^{2}-\la})^{p/q}\cr
\tilde W (y)&= (y+\sqrt{y^{2}-\la ^{p/q)}})^{q/p)}
+(y-\sqrt{y^{2}-\la ^{p/q} })^{q/p}\cr}}
The experience with the ADE and SOS models on fluctuating lattices
\adek
 , \kostov , \dts
 \ allows usto recognize in $X$ and $Y$ the operators
that create the two possible kinds of boundary in the $(p,q)$ model.
 It has been noticed that the ADE models exist in two phases:
dense and dilute. Thus each $(p,q)=(q,p)$ model has two different
statistical realizations. In one of them the boundary has dimension 1,
and in the other a non-classical dimension $p/q$.
The same is true in the case of the $O(n)$ model
\ref\kse{I. Kostov and M. Staudacher, \np B 384 (1992) 459}.
The expansion of the two kinds of macroscopic loops in terms of local
operators contains the two kinds of boundary operators in the model
(see the discussion in section 2.2 of ref. \kse ).

Let us illustrate this
by the  simplest example $p=3,q =2$.
 If we perform the gaussian integration in $Y$,
the resulting matrix model will give the
standard formulation of the $C=0$ theory $(2,3)$ (pure
 gravity).
On the other hand, it is known
\ref\ikvk{V. Kazakov and I. Kostov, {\it unpublished}; I. Kostov,
in Jaca 1988, Proceedings, {\it Non-perturbative aspects of the
standard model}, 295}
\ that this matrix model is equivalent to the $Q=1$ Potts model on
a fluctuating lattice.
If we choose  $g_{1}=h_{1}=0, g_{2}=1, g_{3}=-g$ , then the  temperature
$T$ and the
 ``cosmological constant'' $\beta$ are given by
\eqn\pppot{T=1/\ln h_{2}, \ \ \ \beta = \ln (h_{2}-1) - (2/3)
\ln g}
Integrating over $X$
we find, in the large-$N$ limit, the following one-matrix model \ikvk
\eqn\onmd{\eqalign{
Z&=\int \prod_{i=1}^{N}dy_{i}e^{-\beta V_{eff}(y_{i})}
\prod _{i \ne j}(y_{i}-y_{j})\prod_{i,j}(\sqrt{a-y_{i}}
+\sqrt{a+y_{j}})^{-1/2}\cr
V_{eff}(y)&={1 \over 2}h_{2} \tr Y^{2} +6g\tr Y +{2 \over 3\sqrt{g}}
\tr \sqrt{(a-Y)^{3}}+{1\over N}\tr {1 \over \sqrt{ (a-Y)}}
 \tr \sqrt{(a-Y)}
}}
where the constant $a$ is determined by
\eqn\acafdag{a= {1 \over 4g} - \sqrt{g} {\tr \over N} (a-Y)^{-1/2}}
This model has the same singularity for the free energy as the standard
one-matrix model for pure gravity,   but
 different scal
ing for the loop operators.

Another interesting problem is to find the
 amplitudes  involving two kinds of boundaries
\eqn\dvegr{
W (x,y) = \langle \Tr {1\over x - \hat X }
{1\over x - \hat Y}\rangle
}
whose calculation involves the angular degrees of freedom.

Such amplitudes  have been studied
recently in \ref\ope{V. Kazakov and I. Kostov, \np B 386 (1992) 520}.

\newsec{  Two-loop correlators}

The two-point functions appear in matrix formulation as expectation
values of product of traces;  ${1\over N}\tr X ^m$ corresponds to a
 puncture operator whose action on the string world-sheet leaves
 a hole with a
 boundary of length $m$. Boundaries of any size may be considered,
 with the operator
${1\over N}Tr {1\over x- \tilde X }$, which involves all
 possible punctures, weighted
with the exponential of the length of their boundaries. Such an operator has
 well-defined expectation values, provided the one-dimensional cosmological
 constant $x$
is large enough. The continuum limit is then reached when we let $x$
decrease towards its critical value, so as to allow larger and larger
 boundaries.

There are  three  types of the loop-loop correlation functions
in our model:
\eqn\loopy{\eqalign{
 W_{XY}(x ,y )    =\langle Tr {1\over x
 - \hat X }\ \tr {1\over y   - \hat Y}\rangle _c ,}}
\eqn\loopx{
     W_{XX}(x_1,x_2)     =\langle \tr {1\over x_1
 - \hat X }\ \tr {1\over x_2  - \hat X}\rangle _c }
 \eqn\pooli{    W_{YY}(y_1,y_2)     =\langle \tr {1\over y_1
 - \hat Y }\ \tr {1\over y_2  - \hat Y}\rangle _c }
They may be computed in an explicit way by means of the orthogonal
polynomial technique.
Let us start with the first correlator \loopy .
 In the basis of orthogonal polynomials, we have
\eqn\orpol{\eqalign{
\tilde W(x,y) =
  {1\over N^2}\sum _{i\le N-1,j \ge
N} \langle i|{1\over x - \hat X}|j \rangle \langle j |{1\over y - \hat Y
}|i \rangle      }}
Only the levels located near the Fermi surface $t=\la $  contribute to this
sum, so that, in the planar limit, we can express the former matrix elements
in terms of the functions $X (z),Y (z)$: for instance, $\langle i+
\delta |{1\over x - \hat X}|i \rangle $
is the coefficient of $z^{-\delta}$ in
${1\over x- X(z)}$. Thus we can express the connected two-loop function as
a contour integral and perform the sum over $\delta$
\eqn\ggyg{\eqalign{
             W_{XY}(x,y)         &   = -{1 \over 4\pi ^2} \oint
\oint {dzd\tilde z\over z \tilde z}\sum _{\delta \ge 1}
\delta (z \tilde z)^\delta
 {1\over x - X(z) } {1\over y - Y(\tilde z)}          \cr
&=-{1 \over 4\pi^2} \oint \oint {dzd\tilde z \over(1-z \tilde z)^2}
{1\over x -X(z)}{1\over y- Y(\tilde z)} \cr  }}
where $|z|  <  1, |\tilde z| <1$   for the sum to  converge.

Now,  provided that $x$ and $y$ are big enough,
for the equations $X(z)=x$ and $Y(z)=y$ to have unique solutions
$z(x)$ and $\tilde z(y)$ for $|z|  <  1 , |\tilde z| <1 $,
we calculate the contour integrals by the residua and finally obtain
\eqn\qqaz{W_{XY}(x,y)={z' (x )\tilde z' (y)\over (1-z(x )
\tilde z(y))^2}=-\p_{x}\p_{y} \ln\big[ 1-z(x )\tilde z(y)\big]
}
Note that the double pole at $z= 1 / \tilde z$ does not contribute
since it gives the total derivative under the second integral.

In a similar way we  obtain the other two correlators
\eqn\hhhu{W(x_1,x_2)=-\p_{x_1}\p_{x_2}
\ln\big[{z(x_1)-z(x_2) \over x_1 - x_2 }\big]}
\eqn\hhhuu{W(y_1,y_2)=-\p_{y_1}\p_{y_2}
\ln\big[{ \tilde z(y_1)- \tilde z(y_2) \over y_1 - y_2 }\big]}
We see from these formulae that the knowledge of $z(x)$ and $ \tilde z(y)$
or, in other words, of the one-loop averages, allows the immediate
reconstruction of the corresponding two-loop correlators.

Formulae \qqaz\ - \hhhuu\ work,      in  principle, beyond the two-matrix
model and are applicable to any matrix chain where the orthogonal
polynomial formalism is valid.

In the scaling limit   $z \equiv e^\o \to 1, \o \to 0 $ we  find
in the unitary case $p=q+1$,
using the explicit solution \podo  ,
\eqn\joy{\tilde W_{XY}(X,Y)=
-\p_{X}\p_{Y} \ln\big[  \cosh (\theta) + \cosh (\tau)  \big]
}
\eqn\jox{W_{XX}(X_1,X_2)=
-\p_{X_1}\p_{X_2} \ln\big[ { \cosh (\theta _{1}) - \cosh (\theta _{2}) \over
\cosh (q \theta _{1})- \cosh (q\theta _{2})} \big]
}
and
\eqn\joxx{W_{YY}(Y_1,Y_2)=
-\p_{Y_1}\p_{Y_2} \ln\big[ { \cosh (\tau _{1}) - \cosh (\tau _{2}) \over
\cosh (q \tau _{1})- \cosh (q\tau _{2})} \big]
}
where the following parametrization is used
\eqn\notatt{
 X = 2\mu^{q}\cosh(q\theta), \ \ \
     Y= 2\mu^{p}\cosh(p\tau)}
Expressions \joy - \joxx\ remain true even in the
non-unitary case $p-q > 1$, if the deviation from the critical point is
along the constant $\Lambda$ coupled with the identity operator
(see the discussion at the end of section 7 ).

Let us try to relate the above results with what is known of
the two-loop correlators for strings with discrete target space.

The loop-loop correlator in the one-matrix model with general potential
has been calculated in \make \ (see also the Appendix of \dts ).
 In all critical regimes  ($q=2$, $p$ odd)
 it is given by
\eqn\popi{W(X_{1},X_2)=- {\p \over \p X_1} {\p \over \p X_2}
\log (\sqrt{X_1+\mu^2}+\sqrt{X_2+\mu^2})}
which is identical to the formula \jox\ , which we obtained
 for the $XX$ correlator
with $X=2 \cosh (\theta /2)$.  The
 derivative of this correlator with respect to $\lambda = \mu^{p+1}$
gives the loop-loop-point correlator and factorizes to a product
of two factors
\eqn\iuoo{{ \p \over \p \mu} W(X_{1},X_{2})=
 {1 \over \sqrt{X_1+\mu^2} \sqrt{X_2+\mu^2}}
}
It has been noticed \dts\ that such a factorization is a general property
of all strings with discrete target space. Namely, all multiloop
 amplitudes (three or more loops) depend on the parameters of the
boundaries through a product of factors associated with loops.
In our case the  differentiation in $\mu$ for fixed $X_1,X_2$ gives
\eqn\facx{\eqalign{
\mu {\p \over \p \mu}W_{XX}(X_1,X_2)&={\p \over \p X_1}{\p \over \p X_2}
{1 \over \cosh(\theta_{1})-\cosh(\theta_{2})} \big[
{\sinh[(q-1)\theta_1] \over \sinh (q\theta_1)} -
{\sinh [(p-1)\theta_2]\over \sinh (p\theta_2)}  \big]\cr
&={\p \over \p X_1}{\p \over \p X_2}
\sum_{k=1}^{q-1} {\sinh [(q-k)\theta_1] \over \sinh (q \theta_1)}
{\sinh [(q-k)\theta_2] \over \sinh (q \theta_2) }\cr }}
Similarly, the derivative of the $YY$ correlator decomposes into a
sum of $p-1$ terms.

Comparing \facx\ with the standard expression for the two-loop
correlator (\dts , eq.(4.42)) we can express the operator $P$ in
the $(p,q)$ critical regime of  the
two-matrix model as a linear combination of the order parameters
of the corresponding string theory.
Indeed, the  loop-loop correlator diagonalizes in the
momentum space $k/q , k=1,2,...,q-1,$
Introducing the loop operator  $\Psi_{k}(X)$ with momentum $k/q$
we have
\eqn\loopop{\langle \Psi_{k}(X_1)\Psi_k(X_2)\rangle _c =
{\p \over \p X_1}{\p \over \p X_2}
 {\sinh [(q-k)\theta_1] \over \sinh (q \theta_1)}
{\sinh [(q-k)\theta_2] \over \sinh (q \theta_2) }}
This formula can be obtained from
eq. (4.42) of ref.\dts , or, more easily, from the Laplace image of
 the expression for
the three-loop correlator (eq. (4.58) of ref.\dts ) with one of the loops
shrinked to a point.
Comparing \facx\ and \loopop\ we can conclude that the Heisenberg momentum
$P$ can be written as a linear combination of the loop operators $\Psi_k(X),
k=1,...,q-1$.
In the same way the  operator $Q$ is a linear of the loop operators
$\Psi_{k}(Y), k=1,2,...,p-1$.
Note that the $XX$\  ( $YY$) correlators depend only on $q$ \ ($p$).
This phenomenon occurs only for the two-loop correlator and is
due to the Euler characteristic of the world-sheet being zero.
 The two-loop average, however, contains all the operators; this
is related to the fact that the Euler characteristic of the cylinder
is zero.
As for the mixed correlator $W_{XY}$, its derivative with respect to
$\la$  cannot be written in a factorized form in the non-symmetric case.

All these formulae have an especially nice interpretation in the
$Z_2$-symmetric case of the two-matrix model $U=V$, describing the $(m,m+1)$
unitary models, which we consider in more detail in the next section.

\newsec{Symmetric ($V=U$) realization of the unitary $(m,m+1)$ models}

In this case we can use the equation of motion \ooio\ . The formula
for the multicritical potential is given by \jky\ if one takes there
$p=q=m$.

If one uses only the even potentials (polynomials in the even powers
of the matrices) one has to take, instead of eq.\vxc\ , the following
critical expression for $X_*(z)$:

\eqn\vxs{{\p \over \p z} X_{*}(z)= -{(1-z^2)^{m-1} \over z^2}}
which is the only one satisfying all the conditions of $m$-criticality
and the behaviour for $z \rightarrow 0$. The expression for the
multicritical potential is the same as for the general symmetric case.

Some examples of these potentials are given in the Appendix. Note
that at least the even ($m=2k$) unitary cases provide a
non-perturbatively stable definition of the corresponding rational
string theory in the double scaling limit, since the corresponding
multicritical potential contains the positive highest power coefficient.
As far as we know, this is the simplest possible stabilization of the
double scaling limit of the unitary models of gravity.

The symmetric case provides a direct statistical interpretation of the
two matrix model in terms of the Ising spins on the dynamical random
lattice. In this case the operator $Y$ is the $\Z_{2}$ image of $X$ and
commutes with it. The  loop-loop
correlators  found in the previous section
correspond to the boundary conditions with fixed directions of the
Ising spins: $W_{XX}$ and $W_{YY}$ have the same directions
 of the spin along the two boundaries, whereas $W_{XY}$ has
 opposite directions.
The $\mu$-derivative of the loop-loop correlators
factorizes both for  $W_{XX}=W_{YY}$ and $W_{XY}$:
\eqn\rrpoar{\mu {\p \over \p \mu}W_{XX}(X_{1},X_2)=
{\p \over \p X_1}{\p \over \p X_2}
\sum_{k=1}^{m-1} {\sinh (k\theta_1) \over \sinh (m \theta_1)}
{\sinh (k\theta_2) \over \sinh (m \theta_2) }}
\eqn\rrar{\mu {\p \over \p \mu}W_{XY}(X_1,X_2)=
{\p \over \p X_1}{\p \over \p X_2}
\sum_{k=1}^{m-1}(-)^k {\sinh (k\theta_1) \over \sinh (m \theta_1)}
{\sinh (k\theta _2) \over \sinh (m \theta _2) }}
where
\eqn\ytufdg{X_1=\mu^m\cosh (m\theta_1) ,X_2=\mu^m\cosh (m\theta_2) }

We see that the two-point correlator diagonalizes if we pass to the
 operators imposing symmetric (+) and antisymmetric ($-$)
boundary conditions.


\newsec{Discussion}

We made an attempt to explicitly work out the idea
proposed by M. Douglas about the possibility to represent all models of
$(p,q)$ rational matter interacting with 2D gravity as the multicritical
regimes of the two matrix model, which describes random planar graphs
of specially adjusted statistics with the Ising spins on it.
Although it should be clear to the reader that this purpose has not been
achieved completely in our paper, we made a few steps forward in this
direction, leaving many other important details for the future.

We determined the pairs of $(p,q)$  multicritical potentials, i.e.
the  sets of coupling constants (``temperatures'') rendering the
$(p,q)$ matter critical. By shifting the cosmological coupling
away from its critical value we reconstructed the known scaling solutions
for the one loop averages in these models.

 We found an intriguing relation between  the canonical momentum
 and the loop average in the dispersionless (large $N$) limit
of the generalized KdV formalism.

The two-loop correlators are given in a closed form, which is valid
for arbitrary  potentials.
  Some of our formulae
and the method of calculation apply for any chain of  matrices
admitting orthogonal polynomial representation.
 The scaling form
of loop-loop correlators  is compatible with the previous results
obtained for the RSOS models coupled with gravity:
 in the unitary $(m,m+1)$ model the correlator can
be factorized into $(m-1)$ terms describing the corresponding finite
number of excitations (corresponding to the order operators) propagating
between the macroscopic states (loops).

 One of the most attractive features of the construction
of the  theory of  $(p,q)$ matter
coupled with 2D gravity  {\it via} the
two matrix model, is the explicit  $(p,q)$ versus $(q,p)$
duality. This last property is far from evident in the standard
construction based on the RSOS models on a random lattice.

Among the problems to be addressed further are the following:

1.The calculation of the multi-loop correlators in form as explicit
as was done for the one matrix model.

2. The description  of the double scaling limit and of the topological $1/N$
expansion, using the exact (for any $N$) operator equations \vii .
This becomes especially interesting
 in  view of the possibility to
give within this model a nonperturbative stable definition of the
double scaling limit of   $(2n,2n+1)$ unitary models (note
that the critical  potentials in the Appendix (even case) have a
positive coefficient in front of the highest power for $m=2n$).

3. It would be interesting to compare the $1/N$ expansion in the two matrix
model with the ``surface surgery'' procedure worked out for the ADE models
on random surfaces \dts . In particular, it would be very instructive to
reproduce the open string amplitudes found recently in the framework
of the SOS model \ope . In the two matrix model they must correspond to the
loop amplitudes with the mixed boundary (including the products of both
matrices) in the loop operator.
 Some recent results in this direction were obtained in
\ref\matth{M. Staudacher, {\it Combinatorial solution of the two matrix
model},
Preprint RU-92-64, January 1993}  using the methods of ref.
\ref\alfaro{J.Alfaro,
 preprint CERN-TH-6531/92, July 1992}.

4. A possible application of the formalism worked out here could be
the investigation of the critical regimes of the so-called induced
gauge theory \induce\ in the large N limit.

\bigbreak\bigskip\bigskip\centerline{{\bf Acknowledgements}}\nobreak

We thank   J.Alfaro, D.Boulatov, E.Brezin, M.Douglas,
I. Krichever   and A.Migdal
 for
 discussions.

\appendix{A}{(p,q) Multicritical Potentials for the Two Matrix Model}
\subsec{Nonsymmetric case}
\eqn\multipo{\eqalign{
q=2, p=3: \ \ \ \ \ \ &
V(y)=y - {{{y^2}}\over 2}         \cr
& U(x)=
-x + {{{x^2}}\over 2} + {{{x^3}}\over 3}         \cr
q=3, p=4: \ \ \ \ \ \ & V(y)=
-5\,y + {{5\,{y^2}}\over 2} + {{{y^3}}\over 3}       \cr
& U(x)=
5\,x + {{5\,{x^3}}\over 3} - {{{x^4}}\over 4}          \cr
q=4,p=5 \ \ \ \ \ \ & V(y)=
11\,y - {{11\,{y^2}}\over 2} + {{11\,{y^3}}\over 3} - {{{y^4}}\over 4}
\cr
& U(x)=
-11\,x + {{22\,{x^3}}\over 3} + {{11\,{x^4}}\over 4} + {{{x^5}}\over 5}
\cr
q=2, p=4 \ \ \ \ \ \ & V(y)=
2\,y + {{{y^2}}\over 2}             \cr
& U(x)=
2\,x - {{{x^2}}\over 2} + {{2\,{x^3}}\over 3} + {{{x^4}}\over 4}   \cr
q=3,p=5: \ \ \ \ \ \ & V(y)=
-7\,y - {{7\,{y^2}}\over 2} + {{{y^3}}\over 3}              \cr
& U(x)=
-7\,x + {{7\,{x^3}}\over 3} - {{7\,{x^4}}\over 4} + {{{x^5}}\over 5}
\cr
q=5, p=7: \ \ \ \ \ \ & V(y)=
-23\,y - 23\,{y^2} + {{115\,{y^3}}\over 3} - {{23\,{y^4}}\over 4} +
  {{{y^5}}\over 5}              \cr
&U(x)=
-23\,x + {{115\,{x^3}}\over 3} - {{345\,{x^4}}\over 4} +
  {{161\,{x^5}}\over 5} - {{23\,{x^6}}\over 6} + {{{x^7}}\over 7}
\cr
}}
\subsec{Symmetric case $U=V=V_m$ describing
 the $(m,m+1)$-unitary models $p=q=m$}
\eqn\ai{\eqalign{
& V_3 (x) =
-3\,x - {{3\,{x^2}}\over 2} + {{{x^3}}\over 3}           \cr
& V_4(x)=
8\,x + 2\,{x^2} + {{8\,{x^3}}\over 3} + {{{x^4}}\over 4}       \cr
& V_5(x)=
-15\,x - {{5\,{x^2}}\over 2} + 15\,{x^3} - {{15\,{x^4}}\over 4} +
  {{{x^5}}\over 5}           \cr
& V_6(x)=
24\,x + 3\,{x^2} + {{224\,{x^3}}\over 3} + 39\,{x^4} + {{24\,{x^5}}\over 5} +
  {{{x^6}}\over 6}         \cr
}}
\subsec{Even Symmetric Potentials $V_m$ for $(m,m+1)$ models}

\eqn\mulpot{\eqalign{
 & V_{3}(x) =
2\ {x^2} - {{{x^4}}\over {12}}                       \cr
  & V_{4}(x)=
11\,{x^2} - {x^4} + {{{x^6}}\over {30}}              \cr
  & V_{5}(x)=
{{452\,{x^2}}\over 5} - 13\,{x^4} + {{4\,{x^6}}\over 5} - {{{x^8}}\over {56}}
                                                     \cr
  & V_{6}(x)=
{{64475\,{x^2}}\over {63}} - {{604\,{x^4}}\over 3} + {{158\,{x^6}}\over 9} -
  {{5\,{x^8}}\over 7} + {{{x^{10}}}\over {90}}       \cr
}}
%
%

 \listrefs
\bye